\documentclass[12pt]{article}
\usepackage{epsfig}

\newcommand{\mysection}{\setcounter{equation}{0}\section}

\def\beq{\begin{equation}}
\def\eeq{\end{equation}}
\def\beqa{\begin{eqnarray}}
\def\eeqa{\end{eqnarray}}
 
\newlength{\dinwidth} \newlength{\dinmargin}
\begin{document}

\begin{center}
{\Large \bf NNLL threshold resummation for top-pair and single-top production}
\end{center}
\vspace{2mm}
\begin{center}
{\large Nikolaos Kidonakis}\\
\vspace{2mm}
{\it Kennesaw State University,  Physics \#1202,\\
1000 Chastain Rd., Kennesaw, GA 30144-5591, USA}
\end{center}
 
\begin{abstract}
I discuss threshold resummation at NNLL accuracy in the standard moment-space approach in 
perturbative QCD for top-pair and single-top production. For top quark pair production 
I present new approximate NNLO results for the total cross section and for the top quark transverse momentum 
and rapidity distributions at 8 TeV LHC energy. I discuss the accuracy of the soft-gluon approximation 
and show that the NLO and NNLO approximate results from resummation 
are practically indistinguishable from exact NLO and partial NNLO results. For 
single top production I present new approximate NNLO results for the total cross sections in all three channels at the LHC 
and also for the top quark transverse momentum distributions in $t$-channel 
production and in top-quark associated production with a $W$ boson. 
For both $t{\bar t}$ and single-top production the agreement of theoretical results with LHC and Tevatron data is excellent. 
\end{abstract}
 
\mysection{Introduction}

Threshold resummation is important for increasing the theoretical accuracy of top quark total and 
differential cross sections at hadron colliders. 
The top quark is the heaviest known elementary particle and its study is of unique importance both theoretically 
and in experiments at the LHC and previously at the Tevatron.
The top quark cross section is well-known to receive large 
contributions from soft-gluon emission near partonic threshold. 
Resummation of soft-gluon corrections at next-to-leading-logarithm (NLL) accuracy \cite{KS1,KS2} 
and beyond depends on the color structure of the hard scattering.

Soft-gluon resummation was developed recently at next-to-next-to-leading-logarithm (NNLL) 
accuracy in various approaches for $t{\bar t}$ differential \cite{AFNPY,ttbar,AFNPY2,ttbary} 
and total \cite{AFNPY,ttbar,AFNPY2,HATHOR,BFKS} cross sections. 
It is very important to note that although all these approaches are formally NNLL, they are quite different 
from each other both theoretically and numerically,
and  ``NNLL'' can mean different things since the variables used for the threshold logarithms, the formalism employed, 
and the practical choices made in its implementation vary widely.
Differences between the approaches include performing the resummation for the double-differential cross section 
in single-particle-inclusive (1PI) \cite{ttbar,AFNPY2,ttbary} and/or pair-invariant-mass (PIM) \cite{AFNPY} kinematics
versus doing the resummation solely for the total cross section using production threshold \cite{HATHOR,BFKS}; 
using moment-space resummation in perturbative QCD \cite{ttbar,ttbary,HATHOR} versus using 
Soft-Collinear Effective Theory (SCET) \cite{AFNPY,AFNPY2,BFKS}; choices for the analytical and numerical implementation of 
the expressions and for damping factors away from threshold; and keeping subleading terms of various origins.
A review explaining in detail many of the differences can be found in \cite{NKBP}. 
In the double-differential approach the resummation is sensitive to the kinematical invariants of the partonic process; 
this sensitivity is lost in the total-cross-section-only approach where the production threshold (also known as absolute threshold) 
that is used is simply a special case of the more general partonic threshold employed in 1PI and PIM differential kinematics 
(see discussion in \cite{NKBP}). 
We note that the formalism presented here is the only calculation using the moment-space  
perturbative-QCD NNLL resummation for the double-differential cross section in 
1PI kinematics, $d\sigma/dp_T dY$,
to calculate approximate NNLO total cross sections and transverse momentum, $p_T$, 
and rapidity, $Y$, distributions for both top-pair 
production \cite{ttbar,ttbary} and single-top production \cite{s-ch,tW,t-ch}. 

In the next section we briefly describe our resummation formalism. In Section 3 we present results for top pair production, 
including the total cross section, the top quark transverse momentum distribution, and the top quark rapidity distribution. 
New results for 8 TeV LHC energy are presented and compared with results at 7 TeV. A discussion of the excellent accuracy of the 
soft-gluon approximation at NLO and NNLO is also presented. In Section 4 we discuss new results for single top production 
in the $t$, $s$, and $tW$ channels. New results for the total cross sections at 8 TeV LHC energy are shown together with 
(updated) results at 7 TeV. Furthermore new theoretical calculations and results for the top (and antitop) transverse momentum distributions in 
$t$-channel production are presented at LHC and Tevatron energies. A new calculation for the top quark transverse momentum 
distribution in $tW$ production at the LHC is also presented. We conclude in Section 5.

\mysection{Resummation at NNLL}

We begin with a brief description of our formalism for NNLL resummation 
of soft-gluon corrections.
We consider partonic processes of the form
\beq
f_{1}(p_1)\, + \, f_{2}\, (p_2) \rightarrow t(p)\, + \, X \, ,
\eeq
where $f_1$ and $f_2$ represent partons (quarks or gluons),
$t$ represents the top quark, and $X$ represents
additional final-state particles. 
The partonic cross section explicitly involves the kinematical
invariants  $s=(p_1+p_2)^2$, $t_1=(p_1-p)^2-m_t^2$, $u_1=(p_2-p)^2-m_t^2$, with $m_t$ the top quark mass, 
as well as factorization scale $\mu_F$ and the renormalization scale $\mu_R$. 
The physical cross section is in principle independent of $\mu_F$ and $\mu_R$, but
a dependence appears when we truncate the infinite perturbative series at finite order.

Near partonic threshold for the production of the top-quark final state,
the cross section receives logarithmic contributions that arise from incomplete
cancellations between virtual terms and terms from soft-gluon emission.
These contributions are of the form $[\ln^l(s_4/m_t^2)/s_4]_+$
where $s_4=s+t_1+u_1$ measures distance from partonic threshold.
Soft-gluon resummation depends critically on the color structure of the
partonic process as well as its kinematics.

The resummation of threshold logarithms is performed in moment space, it is based on the 
factorization properties of the cross section, and it employs the 
eikonal approximation for describing the emission
of soft gluons from partons in the hard scattering. 
By taking moments, logarithms of $s_4$ produce 
powers of $\ln N$, with $N$ the moment variable. The resummation was first performed at NLL accuracy in \cite{KS1,KS2} 
and at NNLL accuracy in \cite{ttbar}.

\begin{figure}
\begin{center}
\includegraphics[width=7cm]{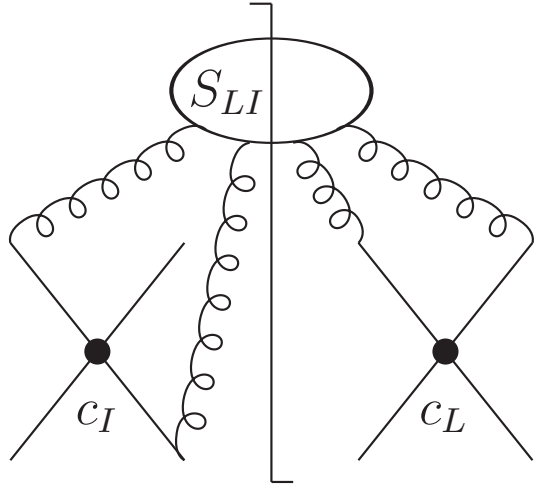}
\caption{The soft-gluon matrix $S_{LI}$. The eikonal lines connect with color tensors $c_I$ and $c_L$.}
\label{eikonal}
\end{center}
\end{figure}

We factorize the moments of the partonic cross section in dimensional regularization
as \cite{KS1,KS2}
\beqa
&&{\hat \sigma}_{f_1 f_2\rightarrow tX}(N,\epsilon)
=\psi_{f_1/f_1} \left(N,\mu_F,\epsilon \right) \;
\psi_{f_2/f_2} \left(N,\mu_F,\epsilon \right)  
\nonumber \\ && \hspace{-10mm} \times \; 
H_{IL}^{f_1 f_2\rightarrow tX} \left(\alpha_s(\mu_R)\right)\; 
S_{LI}^{f_1 f_2 \rightarrow tX} 
\left({m_t\over N \mu_F },\alpha_s(\mu_R) \right)\;
\prod_j  J_j\left (N,\mu_F,\epsilon \right)  
+{\cal O}(1/N) \, ,
\label{sigp}
\eeqa 
where $\psi$ are center-of-mass distributions for the incoming partons, 
$H_{IL}$ is the $N$-independent hard matrix in the space of color exchanges 
(with color indices $I$, $L$), 
$S_{LI}$ is the soft matrix, and $J$ are functions for massless partons in the final state. 
More details about the definitions of these functions and the construction of 
the eikonal cross section can be found in \cite{KS2}.

The hard-scattering matrix involves contributions from the 
amplitude of the process and the complex conjugate of the amplitude,
$H_{IL}=h_L^*\, h_I$.
The soft function $S_{LI}$ represents the emission of noncollinear soft gluons 
from the partons in the scattering. 
The color tensors of the hard scattering connect
together the eikonal lines to which soft gluons couple (see Fig. 1).

The $N$-dependence of the soft matrix $S_{LI}$ can be resummed by renormalization group
analysis. $S_{LI}$ satisfies the renormalization group equation \cite{KS1,KS2}
\begin{equation}
\left(\mu {\partial \over \partial \mu}
+\beta(g_s){\partial \over \partial g_s}\right)\,S_{LI}
=-(\Gamma^\dagger_S)_{LB}S_{BI}-S_{LA}(\Gamma_S)_{AI}\, ,
\label{RGE}
\end{equation}
where $\beta$ is the QCD beta function and $g_s^2=4\pi\alpha_s$.
$\Gamma_S$ is the soft anomalous dimension matrix 
and it is calculated in the eikonal approximation 
by explicit renormalization of the soft function.

The exponentiation of logarithms of $N$ in the functions $\psi$ and $J$ 
in the factorized cross section, 
together with the solution of the renormalization group equation (\ref{RGE}), 
provide us with the complete expression for the resummed 
(double-differential) partonic cross section in moment space 
\beqa
{\hat{\sigma}}_{res}(N) &=&   
\exp\left[ \sum_i E^{f_i}(N_i)\right] \; 
\exp\left[ \sum_j {E'}^{f_j}(N')\right] \; 
\exp \left[\sum_i 2\int_{\mu_F}^{\sqrt{s}} \frac{d\mu}{\mu}\;
\gamma_{f_i/f_i}\left(\alpha_s(N_i,\mu)\right)\right] \;
\nonumber\\ && \hspace{-10mm} \times \,
{\rm Tr} \left \{H^{f_1 f_2 \rightarrow tX}\left(\alpha_s(\sqrt{s})\right) \;
\exp \left[\int_{\sqrt{s}}^{{\sqrt{s}}/{\tilde N'}} 
\frac{d\mu}{\mu} \;
\Gamma_S^{\dagger \, f_1 f_2 \rightarrow tX}\left(\alpha_s(\mu)\right)\right] 
\right. 
\nonumber\\ && \hspace{-10mm} \left. \times \,
S^{f_1 f_2 \rightarrow tX} \left(\alpha_s\left(\frac{\sqrt{s}}
{\tilde N'}\right) \right) 
\exp \left[\int_{\sqrt{s}}^{{\sqrt{s}}/{\tilde N'}} 
\frac{d\mu}{\mu}\; \Gamma_S^{f_1 f_2 \rightarrow tX}
\left(\alpha_s(\mu)\right)\right] \right\} \, .
\label{resHS}
\eeqa

The first and second exponents in Eq. (\ref{resHS}) 
control collinear and soft gluon emission \cite{GS,CT} from incoming and outgoing partons, respectively, 
while the third exponent controls the factorization scale dependence of the cross section. 
Explicit expressions for these exponents can be found in Ref. \cite{NKdpf11}.
The evolution of the soft gluon function is controlled by the soft anomalous dimension matrix
$\Gamma_S$ via the solution to Eq. (\ref{RGE}).

The ultraviolet poles in loop diagrams involving eikonal lines 
play a direct role in the renormalization
group evolution equations that are used in the calculation 
of the soft anomalous dimensions \cite{KS1,KS2,NK2l}. 
The soft anomalous dimension matrices in our formalism have been calculated at two loops for the partonic processes in 
top-pair production in \cite{ttbar,NK2l} and for single-top production in \cite{s-ch,tW,t-ch}.

\mysection{Top-pair production}

The threshold resummation formalism discussed here has been used to calculate top-pair production at the Tevatron and the 
LHC \cite{ttbar,ttbary}. We expand the NNLL resummed cross section to NNLO. We add the NNLO soft-gluon corrections from this
expansion to the exact NLO \cite{NLO1,NLO2,NLO3} expressions and denote the result as approximate NNLO.
Here we present new results for the current 8 TeV LHC energy for the total cross section and the top quark 
transverse momentum and rapidity distributions. The factorization and renormalization 
scales are set equal to each other and denoted by $\mu$. Throughout we use the 
MSTW2008 \cite{MSTW} NNLO parton distribution functions (pdf).

\begin{figure}
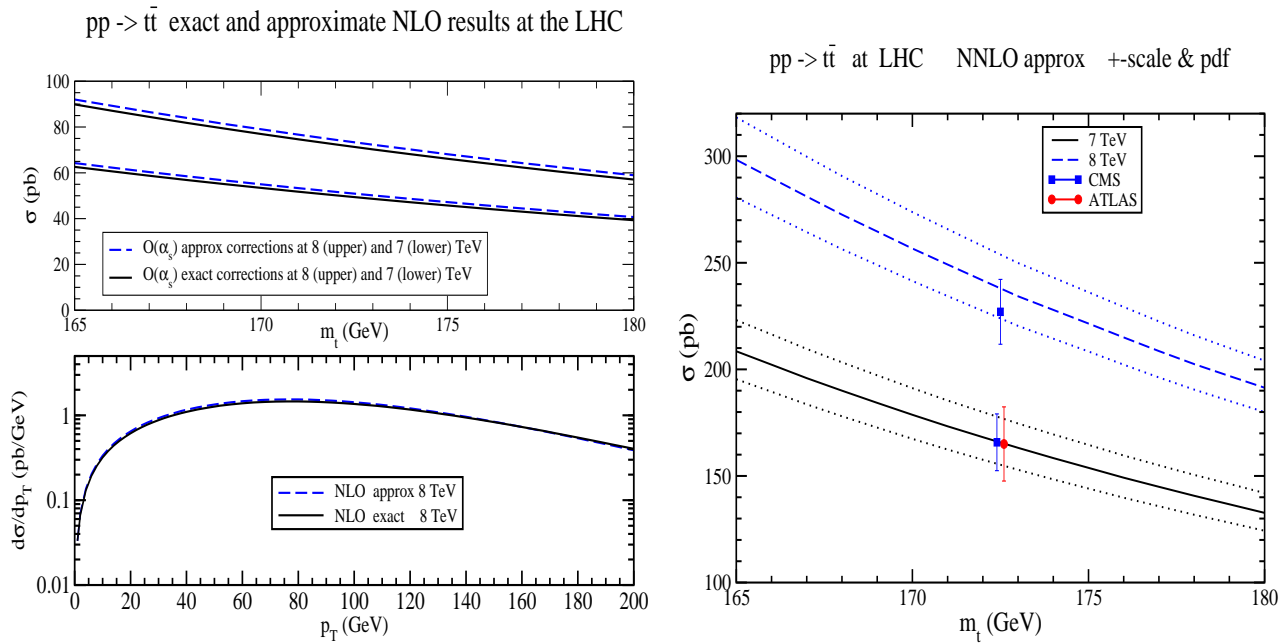

\begin{center}
\includegraphics[width=85mm,height=85mm]{sigptcorr8lhcmplot.eps}
\hspace{1mm}
\includegraphics[width=8cm,height=80mm]{ttblhcplot.eps}
\caption{Comparison of exact and approximate NLO results for the total top-pair cross section and top-quark $p_T$ distribution (left); NNLO approximate top-pair total cross section compared with data at the LHC (right).}
\label{toplhc}
\end{center}
\end{figure}

We begin by examining the validity and numerical accuracy of the threshold soft-gluon 
approximation by comparing exact NLO and approximate NLO results (with $\mu=m_t$) for the 
total $t{\bar t}$ cross section and the top quark $p_T$ distribution. The comparison is 
shown in the left two plots of Fig. \ref{toplhc}. The top left plot displays the exact 
and approximate NLO corrections, i.e. ${\cal O}(\alpha_s)$ corrections, for the total 
top-pair cross section at 7 and 8 TeV LHC energy. We note that the approximation is excellent 
with only around 2 to 3\% difference between approximate and exact corrections. The 
approximate results include only $q{\bar q}$ and $gg$ channels (for which resummation is 
performed) while the exact results also include the $qg$ and ${\bar q}g$ channels 
(the contribution from the latter two channels is quite small). In fact if only the sum of 
the $q{\bar q}$ and $gg$ channels are included in the exact result the difference 
between approximate and exact corrections is only 1 to 2\%. This very good agreement between 
exact and approximate NLO corrections also holds separately for the $q{\bar q}$ and $gg$ 
channels. If one compares the total NLO cross section 
(i.e. the sum of the leading-order cross section plus the NLO corrections) 
the difference between approximate and exact results is entirely negligible, 
well below 1\%. This excellent agreement between exact and approximate NLO results is true 
not only for the total cross section but also for differential distributions. The bottom left 
plot of Fig. \ref{toplhc} shows that the exact NLO transverse momentum distribution of the top quark is again 
almost identical to the approximate NLO result (the results are for 8 TeV LHC energy and 
$m_t=173$ GeV). We note that there are some choices to be made in the analytical structure 
and numerical implementation of the calculation, all strictly equivalent at partonic threshold
but not entirely equivalent away from it, and different choices can give somewhat different 
results. The agreement between exact and approximate results shown here proves the validity, 
accuracy, and importance of our approach and the appropriateness of our choices, and it is 
the motivation for studying higher-order soft-gluon corrections in the same approach 
and with the same choices. As we will discuss shortly, 
recent partial NNLO results are also in excellent agreement with the 
approximate NNLO results from our formalism, thus further proving the advantages and 
accuracy of our approach.  

In the right plot of Fig. \ref{toplhc} we show the NNLO approximate total top-pair production 
cross section at the LHC at 8 TeV energy, and for comparison also at 7 TeV energy, as a 
function of top quark mass. Here we have set $\mu=m_t$ for our central results. 
The uncertainties from scale variation and from the pdf are added in quadrature.
The results at 7 TeV are compared with ATLAS and CMS data \cite{ATLAStt7,CMStt7}. 
The 8 TeV results are compared with recent CMS data \cite{CMStt8}.  
Excellent agreement is found between theory and experiment for both energies. 

The new result for the current 8 TeV LHC energy with a top quark mass $m_t=173$ GeV and using MSTW2008 
NNLO pdf is 
\beqa
\sigma^{\rm NNLOapprox}_{t{\bar t}}(m_t=173\, {\rm GeV}, \, 8\, {\rm TeV})&=&234 {}^{+10}_{-7} \pm 12 \; {\rm pb} \, ,  
\eeqa
where the first uncertainty is from scale variation $m_t /2 < \mu < 2m_t$ and 
the second is from the pdf at 90\% C.L. We note that independent variation of $\mu_F$ and $\mu_R$ does not increase 
the scale uncertainty at LHC energies. At 7 TeV the corresponding result is $163 {}^{+7}_{-5} \pm 9$ pb.

At the Tevatron the total top-pair cross section is $7.08 {}^{+0.00}_{-0.24} {}^{+0.36}_{-0.27}$ pb \cite{ttbar}, 
in very good agreement with recent results from CDF \cite{CDFtt} and D0 \cite{D0tt}. 
Independent variation of $\mu_F$ and $\mu_R$ changes the upper scale error at the Tevatron from $+0.00$ to $+0.20$.
A lot of theoretical work by many groups over the last few years (see \cite{NKBP} for complete 
references) has made possible the calculation of exact NNLO corrections. Recently there have 
been numerical results of the exact NNLO contribution from the fermionic channels to the total 
top-pair cross section \cite{BCM}. Since the $q{\bar q}$ 
channel is dominant at the Tevatron this is supposed to be a good approximation to the still unknown complete 
NNLO at that collider. We note that our approximate NNLO results at the Tevatron are virtually indistinguishable 
from the results in \cite{BCM} (the difference between $7.08 {}^{+0.20}_{-0.24}$ \cite{ttbar} 
versus $7.07 {}^{+0.20}_{-0.31}$ pb \cite{BCM} 
for $m_t=173$ GeV is at the per mille level with very similar scale uncertainty from independent $\mu_F$ and $\mu_R$ variation; 
in fact the scale dependence of our result is slightly smaller than that of \cite{BCM}), 
indicating that future complete NNLO results will make very little, if any, practical difference for Tevatron and probably LHC 
measurements of the $t{\bar t}$ cross section. We also emphasize that the comparison between exact and approximate 
results from resummation in \cite{BCM} is quite distinct from ours since different resummation formalisms are used. 

\begin{figure}
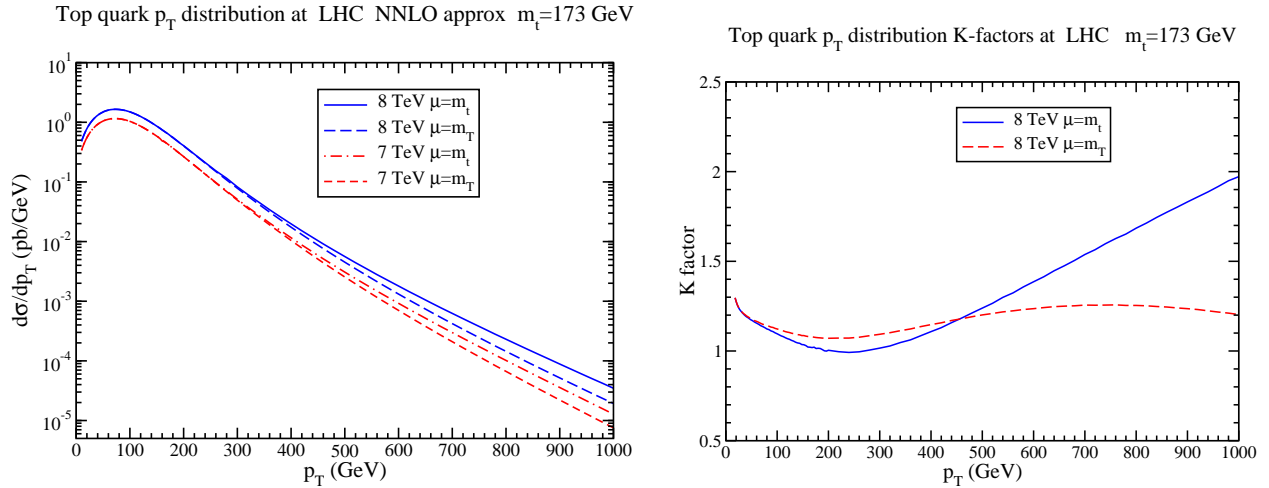

\begin{center}
\includegraphics[width=83mm]{ptlhcmTmplot.eps}
\hspace{3mm}
\includegraphics[width=77mm]{Kptlhcplot.eps}
\caption{NNLO approximate top-quark $p_T$ distributions (left) and related $K$-factors (right) at the LHC.}
\label{ptlhc}
\end{center}
\end{figure}

The fact that corrections beyond the soft-gluon approximation make very little difference was 
actually inferred from the study in \cite{NKRV}. The calculation there was done in both 
1PI and PIM kinematics and it was shown 
that when terms beyond NLL are included, the difference between the results is reduced 
and vanishes at partonic threshold. 
This indicated that further terms would make very little difference, and thus the fact that 
the complete NNLL terms and now the exact NNLO terms make very little difference was largely 
expected. This was also discussed in \cite{NKRV2} and \cite{ttbar}. The more recent calculations  
at NNLL and NNLO prove the correctness of our earlier arguments and the robustness of our results.

No NNLO differential distributions are known at present but given the agreement of the exact and approximate NLO 
distributions, the agreement between the approximate and partially exact NNLO total cross sections, and the lessons 
learned from the comparison of approximate NLO and NNLO results calculated in 1PI and PIM kinematics, we expect 
that approximate NNLO should be sufficient for all practical purposes and exact NNLO results will make no 
significant difference to the distributions.

We also note that once the complete NNLO results are known for either the total cross section or differential 
distributions, the next natural step will be to add the approximate NNNLO corrections. In fact, a first study 
of those was performed in \cite{NKNNNLO} for the $q{\bar q}$ channel at the Tevatron and it was found that 
the NNNLO corrections are small although not insignificant.

\begin{figure}
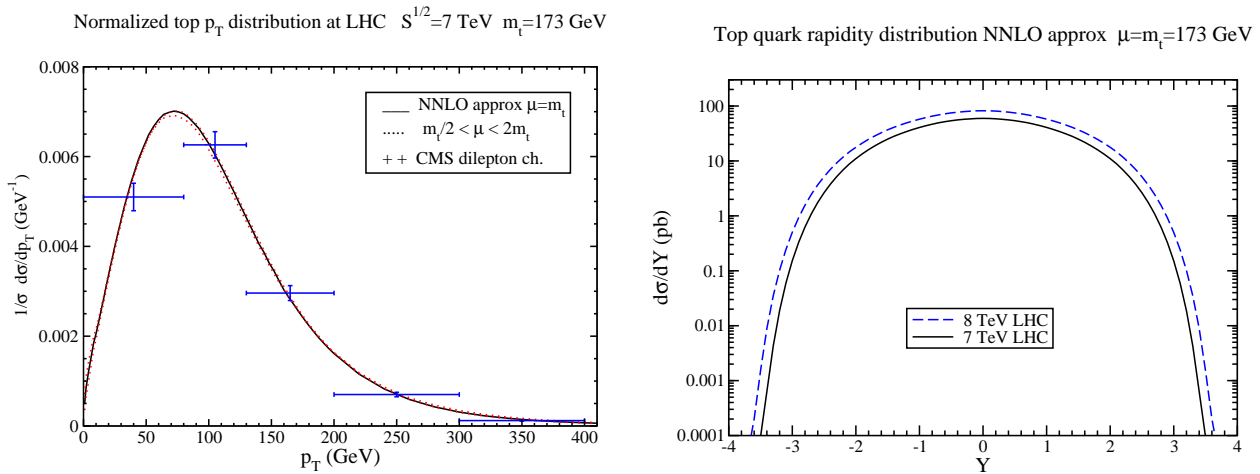

\begin{center}
\includegraphics[width=8cm]{pt7lhcnormCMSdileptplot.eps}
\hspace{3mm}
\includegraphics[width=8cm]{ylogplot.eps}
\caption{NNLO approximate top-quark normalized $p_T$ distribution compared with CMS data (left) and top-quark rapidity distributions (right) at the LHC.}
\label{ptylhc}
\end{center}
\end{figure}

A major strength of our double-differential calculation is that it can be used to calculate differential 
distributions. Of particular significance is the study of transverse momentum and rapidity distributions.
In the left plot of Fig. \ref{ptlhc} we show the top-quark transverse momentum distribution, $d\sigma/dp_T$, 
at the LHC. New results for 8 TeV LHC energy are plotted and compared with results at 7 TeV. The  
approximate NNLO top quark $p_T$ distributions are shown with a top quark mass $m_t=173$ GeV and the 
scale choices $\mu=m_t$ and $\mu=m_T$, where the transverse mass $m_T$ is defined by $m_T=(p_T^2+m_t^2)^{1/2}$. 
In the right plot of Fig. \ref{ptlhc} we show the $K$-factors, i.e. the ratios of NNLO approximate / NLO results for the top-quark $p_T$ distribution at 8 TeV LHC energy with scale choices $\mu=m_t$ and $\mu=m_T$. We observe that the $K$-factor for $\mu=m_t$ rises significantly at large $p_T$ as contrasted to that with $\mu=m_T$.

In the left plot of Fig. \ref{ptylhc} we present the normalized top-quark $p_T$ distribution, $(1/\sigma) \, d\sigma/dp_T$, at 7 TeV LHC energy and compare it with recent data from CMS in the dilepton channel. We note the very good description of the CMS data by the theoretical predictions. Similar agreement is also found with CMS $l$+jets data \cite{CMSpty} as shown in \cite{NKckm12}. 

In the right plot of Fig. \ref{ptylhc} we show new results for the top quark rapidity distribution, $d\sigma/dY$,
at the LHC at 8 TeV energy and, for comparison, at 7 TeV as well. The approximate NNLO top quark rapidity 
distributions are shown with a top quark mass $m_t=173$ GeV and the scale choice $\mu=m_t$. 
The NNLO soft gluon corrections enhance the distribution without a significant change in shape relative to NLO. 
Our results are also in excellent agreement with CMS data \cite{CMSpty} 
as shown in \cite{NKckm12}.

We also note that results for $p_T$ distributions and for rapidity distributions and the forward-backward asymmetry 
at the Tevatron have been presented in \cite{ttbar} and \cite{ttbary} respectively, and a comparison with D0 data \cite{D0pt} shows excellent agreement \cite{NKBP} of our $p_T$ distribution calculation with Tevatron results. 

\mysection{Single-top production}

We continue with single top quark production at the LHC and the Tevatron. For the total cross 
sections we update results at 7 TeV \cite{s-ch,tW,t-ch} and present new results at 8 TeV LHC 
energy for $t$-channel, $s$-channel, and associated $tW$ production. We also present new 
theoretical calculations and results for the single-top and single-antitop transverse 
momentum distributions in $t$-channel production at the Tevatron and at 7 and 8 TeV energy 
at the LHC and in $tW$ production at the LHC. As for top-pair production, we note that the 
agreement between exact NLO \cite{HLPSW,Zhu}
and approximate NLO cross sections for all three single-top 
channels is very good, which again serves as the motivation to derive approximate NNLO 
results. 

At leading order, the partonic processes for $t$-channel production are 
$qb \rightarrow q' t$ and ${\bar q} b \rightarrow {\bar q}' t$; 
for $s$-channel production, $q{\bar q}' \rightarrow {\bar b} t$; and for 
associated $tW$ production, $bg \rightarrow tW^-$. At the Tevatron the cross sections for single top production in all three channels are identical to those for single antitop. However at the LHC the $t$-channel and $s$-channel single top cross sections are larger than those for antitop.

\begin{table}[h]
\begin{center}
  \begin{tabular}{c|c|c|c}
LHC 7 TeV & $\sigma(t)$ (pb) & $\sigma({\bar t})$ (pb) & 
$\sigma(t)+\sigma({\bar t})$ (pb) \\ \hline
$t$-channel  & 
$43.0 {}^{+1.6}_{-0.2} \pm 0.8$ &
$22.9 \pm 0.5 {}^{+0.7}_{-0.9}$ &
$65.9 {}^{+2.1}_{-0.7} {}^{+1.5}_{-1.7}$
\\
$s$-channel  & 
$3.14 \pm 0.06 {}^{+0.12}_{-0.10}$ &
$1.42 \pm 0.01 {}^{+0.06}_{-0.07}$ &
$4.56 \pm 0.07 {}^{+0.18}_{-0.17}$
\\ 
$tW$ &
$7.8 \pm 0.2 {}^{+0.5}_{-0.6}$ &
$7.8 \pm 0.2 {}^{+0.5}_{-0.6}$ &
$15.6 \pm 0.4 \pm 1.1$ 
\end{tabular}
\end{center}
\vspace{-2mm}
\caption{\label{tab:singletop1} Results for single-top, single-antitop, and 
combined approximate NNLO cross sections at 7 TeV LHC energy with $m_t=173$ GeV
in pb.}
\end{table}

\begin{table}[h]
\begin{center}
  \begin{tabular}{c|c|c|c}
LHC 8 TeV & $\sigma(t)$ (pb) & $\sigma({\bar t})$ (pb) & 
$\sigma(t)+\sigma({\bar t})$ (pb) \\ \hline
$t$-channel  & 
$56.4 {}^{+2.1}_{-0.3} \pm 1.1$  &
$30.7 \pm 0.7 {}^{+0.9}_{-1.1}$  & 
$87.2 {}^{+2.8}_{-1.0} {}^{+2.0}_{-2.2}$  
\\
$s$-channel  & 
$3.79 \pm 0.07 \pm 0.13$  &
$1.76 \pm 0.01 \pm 0.08$  & 
$5.55 \pm 0.08 \pm 0.21$ 
\\ 
$tW$ &
$11.1 \pm 0.3 \pm 0.7$  &
$11.1 \pm 0.3 \pm 0.7$  &
$22.2 \pm 0.6 \pm 1.4$
\end{tabular}
\end{center}
\vspace{-2mm}
\caption{\label{tab:singletop2} Results for single-top, single-antitop, 
and combined approximate NNLO cross sections at 8 TeV LHC energy with 
$m_t=173$ GeV in pb.}
\end{table}

In Tables 1 and 2 we present for 7 and 8 TeV LHC energy, respectively, the total 
NNLO approximate cross sections, derived from NNLL resummation, for single top production 
in the $t$ channel, $s$ channel, and in $tW$ production. Separate results are given for single 
top, single antitop, and the combined sum. In each case, the central result is for $\mu=m_t$, the first uncertainty is from scale variation 
$m_t /2 < \mu < 2m_t$, and the second uncertainty is from the MSTW2008 NNLO 
pdf sets at 90\% CL \cite{MSTW}.
The $t$ channel has the largest cross section, followed by $tW$ production, 
with the $s$ channel numerically the smallest.
Our theoretical results for the Tevatron can be found in \cite{s-ch,t-ch} and 
they are in good agreement with recent Tevatron data; the total cross section that we find for the sum of $t$ plus $s$ channels is $3.16^{+0.18}_{-0.19}$ pb 
for $m_t=172.5$ GeV which agrees well with
the results of $3.04^{+0.57}_{-0.53}$ pb from CDF \cite{CDFts} and 
$3.43^{+0.73}_{-0.74}$ pb from D0 \cite{D0ts}.

\begin{figure}
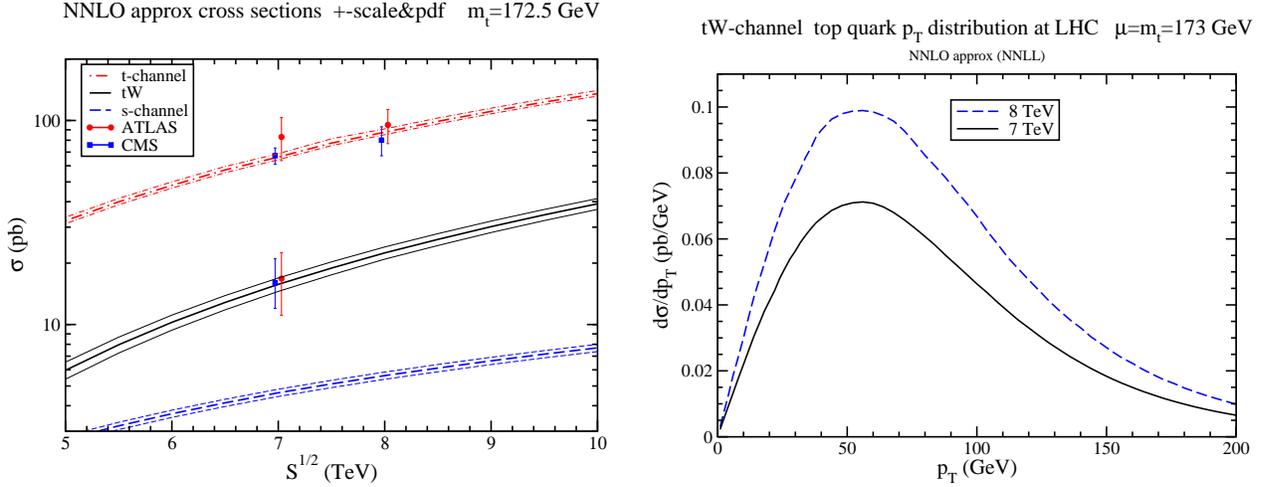

\begin{center}
\includegraphics[width=8cm]{tch-tW-sch-comboplot.eps}
\hspace{3mm}
\includegraphics[width=8cm]{pttWlhcplot.eps}
\caption{Total $\sigma(t)+\sigma({\bar t})$ approximate NNLO cross sections in all three single-top channels compared with data from the LHC (left); 
$tW$-channel top-quark $p_T$ distribution at the LHC (right).}
\label{singletop}
\end{center}
\end{figure}

In the left plot of Fig. \ref{singletop} we show the total cross sections for the three single-top channels versus the LHC energy, with the central result for $\mu=m_t$, and the scale and pdf uncertainties added in quadrature. 
The $t$-channel results are in very good agreement with LHC data from both ATLAS and CMS at 7 TeV \cite{ATLAStch7,CMStch7} and 8 TeV \cite{ATLAStch8,CMStch8}. 
The $tW$ results at 7 TeV are also in excellent agreement with data from ATLAS \cite{ATLAStW} and CMS \cite{CMStW}. At present there is only a limit on the $s$-channel cross section 
$<26$ pb at 95\% CL \cite{ATLASsch}.

We also note that the ratio of the single-top and single-antitop cross sections in the $t$-channel 
has been measured by ATLAS \cite{ATLASratio} at 7 TeV LHC energy. The ATLAS result of $1.81^{+0.23}_{-0.22}$ is in excellent agreement with our calculation of $1.88^{+0.11}_{-0.09}$.

In the right plot of Fig. \ref{singletop} we show results from a new calculation for the top quark transverse momentum distribution in $tW$ production at the LHC at both 7 and 8 TeV collider energy.

\begin{figure}
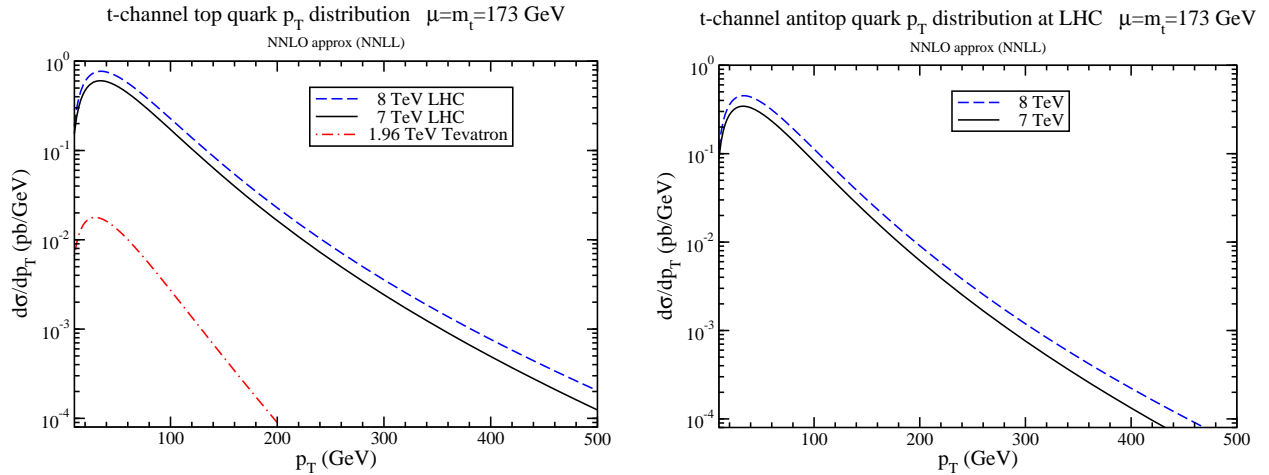

\begin{center}
\includegraphics[width=8cm]{pttchtopcomboplot.eps}
\hspace{3mm}
\includegraphics[width=8cm]{pttchantitoplhclogplot.eps}
\caption{$t$-channel approximate NNLO top-quark (left) and antitop (right) $p_T$ distributions
at the LHC and the Tevatron.}
\label{tchpt}
\end{center}
\end{figure}

In Fig. \ref{tchpt} we show new results for the top quark transverse momentum distribution in $t$-channel single top production at the LHC and the Tevatron (left plot) and the antitop quark transverse momentum distribution in $t$-channel production at the LHC (right plot). 

\mysection{Conclusion}

The threshold resummation of soft-gluon contributions provides a powerful method to accurately 
calculate top quark total cross sections and differential distributions for both top-pair and 
single-top production. Resummation is performed at NNLL accuracy for the double-differential cross section 
and used to provide approximate NNLO results for total and differential cross sections. 
We have shown the applicability, validity, and accuracy of the approach by explicit 
comparisons to complete NLO and partial NNLO calculations. The comparisons show that currently 
available partial exact NNLO corrections are virtually indistinguishable from the NNLO approximations in 
our formalism (which is distinct from other resummation formalisms) both in the central result and in the 
theoretical uncertainty,  
and they thus indicate that future complete NNLO corrections will very likely have a negligible  
impact on the existing results from our formalism. 
New state-of-the-art results for top-pair and single-top total cross sections and transverse 
momentum and rapidity distributions are presented at LHC and Tevatron energies, with 
particular attention to the 8 TeV LHC energy. All theoretical results, for both $t{\bar t}$ and 
single-top production, and for both total cross sections and differential distributions,  are in excellent 
agreement with recent LHC and Tevatron data.

\section*{Acknowledgements}
  
This material is based upon work supported by the National Science Foundation under Grant No. PHY 1212472.

\end{document}